\documentclass[journal]{IEEEtran}
\usepackage{cite}
\usepackage{graphicx}
\usepackage{times}
\usepackage{latexsym}

\usepackage[mathscr]{eucal}
\usepackage[center]{caption2}
\usepackage{array}
\usepackage{fancyhdr}
\usepackage{amsmath,amssymb}
\usepackage{bm} 
\usepackage{multirow}
\usepackage{algpseudocode}
\usepackage{algorithm}
\usepackage{amsfonts}
\usepackage{threeparttable}

\ifCLASSINFOpdf
\else
\fi

\newcommand{\Diag}{\text{Diag}}

\newcommand{\nn}{\nonumber}

\newcommand{\vect}[1]{{\lowercase{\mathbf{#1}}}}
\newcommand{\mat}[1]{{\uppercase{\mathbf{#1}}}}

\newcommand{\g}{\vect{g}}
\newcommand{\h}{\vect{h}}
\newcommand{\m}{\vect{m}}

\newcommand{\F}{\mat{F}}
\newcommand{\G}{\mat{G}}
\renewcommand{\H}{\mat{H}} 

\newcommand{\K}{\mat{K}}

\newcommand{\Cb}{{\mathbb C}}

\newcommand{\Eb}{{\mathbb E}}

\newcommand{\Rb}{{\mathbb R}}

\newcommand{\Lambdam}{\hbox{\boldmath$\Lambda$}}

\usepackage{xcolor}

\usepackage{subfigure}

\makeatother

\newcommand{\Paragraph}[1]{\smallskip\noindent{\bf #1.}}
\usepackage[group-separator={,}, per-mode=symbol, binary-units]{siunitx}

\graphicspath{{Figure/}}

\begin{document}
\title{RIS-Jamming: Breaking Key Consistency in Channel Reciprocity-based Key Generation}

\author{Guyue Li,~\IEEEmembership{Member,~IEEE}, Paul Staat, Haoyu Li, Markus Heinrichs, Christian Zenger~\IEEEmembership{Member,~IEEE}, Rainer Kronberger,~\IEEEmembership{Member,~IEEE}, Harald Elders-Boll, Christof Paar~\IEEEmembership{Fellow,~IEEE} and Aiqun Hu,~\IEEEmembership{Senior Member,~IEEE}

	\thanks{Guyue Li and Haoyu Li are with the School of Cyber Science and Engineering, Southeast
	University, Nanjing 210096, China (e-mail: guyuelee@seu.edu.cn;haoyuli@seu.edu.cn).}
	\thanks{Paul Staat and Christof Paar are with Max Planck Institute for Security and Privacy, Bochum, Germany (e-mail: paul.staat@mpi-sp.org;christof.paar@mpi-sp.org).}
	\thanks{Markus Heinrichs, Rainer Kronberger, and Harald Elders-Boll are with TH Köln – University of Applied Sciences, Cologne, Germany  (e-mail: markus.heinrichs@th-koeln.de;rainer.kronberger@th-koeln.de;harald.elders-boll@th-koeln.de).}
    \thanks{Christian Zenger is with PHYSEC GmbH, Bochum, Germany and Ruhr University Bochum, Germany  (e-mail: christian.zenger@rub.de).}
	\thanks{Aiqun Hu is with the National Mobile Communications Research Laboratory, Southeast University, Nanjing 210096, China (e-mail: aqhu@seu.edu.cn).}
	\thanks{Guyue Li and Aiqun Hu are also with the Purple Mountain Laboratories for Network and Communication Security, Nanjing 210096, China.
	 }
 }
\maketitle
\begin{abstract}
Channel Reciprocity-based Key Generation (CRKG) exploits reciprocal channel randomness to establish shared secret keys between wireless terminals. 
This new security technique is expected to complement existing cryptographic techniques for secret key distribution of future wireless networks. 
In this paper, we present a new attack, reconfigurable intelligent surface (RIS) jamming, and show that an attacker can prevent legitimate users from agreeing on the same key by deploying a malicious RIS to break channel reciprocity. Specifically, we elaborate on three examples to implement the RIS-jamming attack: Using active nonreciprocal circuits, performing time-varying controls, and reducing the signal-to-noise ratio. The attack effect is then studied by formulating the secret key rate with a relationship to the deployment of RIS. To resist such RIS-jamming attacks, we propose a countermeasure that exploits wideband signals for multipath separation. {The malicious RIS path is  
distinguished from all separated channel paths, and thus the countermeasure is referred to as contaminated path removal-based CRKG (CPR-CRKG). }
We present simulation results, showing that legitimate users under RIS jamming are still able to generate secret keys from the remaining paths. We also experimentally demonstrate the RIS-jamming attack by using commodity Wi-Fi devices in conjunction with a fabricated RIS prototype. In our experiments, we were able to increase the average bit disagreement ratio~(BDR) of raw secret keys by 20\%. Further, we successfully demonstrate the proposed CPR-CRKG countermeasure to tackle RIS jamming in wideband systems as long as the source of randomness and the RIS propagation paths are separable.
\end{abstract}


\section{Introduction} 
\label{sec:introduction}
The steady growth in connectivity, fueled by the advent of electronic-commerce applications and the ubiquity of wireless communications, has led to an unprecedented awareness of the importance of network security in all its guises~\cite{ylianttila20206g}. 
In the realm of wireless security, traditional cryptographic schemes are used to secure communications, i.e., symmetric and public-key schemes. 
However, these schemes involve additional effort for, e.g., a public-key infrastructure, key sharing, or key management. Further, the loss and theft of encryption keys pose additional risks. Finally, especially for asymmetric schemes, the notion of computational security is weakened by the development of efficient algorithms as well as the increase in computational power of modern computers (e.g., quantum computers).

In this context, Channel Reciprocity-based Key Generation (CRKG) has emerged as a promising technique 
that provides quantum-resistant secure key sharing among constrained devices. The technique is developed on the source model of physical-layer security~\cite{maurer1993secret}, where two legitimate terminals observe a common random source that is inaccessible to an eavesdropper. Such a natural shared entropy source exploited by CRKG is the wireless fading channel~\cite{08MOBICOM}.
Suppose that a pair of wireless terminals communicate with each other on the same frequency in a wireless communication environment. The wireless channel between two terminals produces a random mapping between the transmitted and received signals. This mapping, known as channel state information (CSI), changes with time in a manner that is location-specific and reciprocal, i.e., the mapping is essentially the same in both directions. 
Hence, if both terminals possess some means of observing the fading of their mutual channel at approximately the same time, their resulting observations are highly statistically dependent. For example, in time-division duplex (TDD) systems, when the coherence time of the channel is sufficiently long, these mutual channels are approximately identical.  Additionally, this time-varying mapping decorrelates completely over distances of the order of a few wavelengths. 
During the past thirty years, much work has been devoted to theoretical study as well as to prototype implementation of CRKG under various wireless communication protocols~\cite{9149584}.

Since CRKG has been recognized as a promising candidate solution to complement
existing cryptographic techniques for 6G~\cite{9762838}, it further incentivizes the need to study the performance of CRKG under emerging attacks.
For example, the emerging technique of Reconfigurable Intelligent Surface (RIS) creates a new idea to deteriorate the fundamental principle of channel reciprocity, resulting in frequent failures of key agreements. RIS is a kind of artificial surface that consists of a large number of sub-wavelength unit cells with tunable electromagnetic responses~\cite{9140329}. 
It can realize the regulation of electromagnetic waves and thus can change the wireless propagation environment.  Compared to other wireless technologies, the advantage of the RIS is its cost-efficient manufacturing and deployment. That is, a RIS consists of a printed circuit board that does not generate its own wireless signals but merely interacts with ambient signals, hence it is referred to as passive~\cite{9140329}. In the future -- RIS is in discussion to be used as a key technology for 6G networks -- RIS deployments are expected to be found on building facades, indoor walls, furniture, and roadside billboards.
When a RIS participates in the process of CRKG, the effective channel is the superimposition of the direct link and the RIS-induced link, the latter of which is given by the product of two physical channel gains and the RIS gain. The success of CRKG, in this context, is also affected by the reciprocity of the RIS-induced link. In other words, the reciprocity property of the entire channel could be destroyed by deploying a malicious RIS that does not satisfy these conditions. In this paper, we define such kind of attack as~\textit{RIS-jamming attack}, as the reciprocal direct link is jammed by the RIS-induced link intentionally. 
As a RIS is considered to be nearly passive, this attack differs from existing jamming attacks~\cite{MiM2021,17TIFSArsa,18TVTXiao}, thereby urging a new model to study its effect and a reconsideration of the countermeasure.

  
This paper aims to address the security problem of CRKG in wireless communication systems in the presence of a malicious RIS. Please note that the RIS is a rather new technology that is currently emerging in the area of communications engineering. Here, it is mostly employed to improve wireless communications, e.g., to maximize the signal power arriving at a receiver. Our work is one of the few to investigate the adversarial potential of RIS -- an important field of study, as the RIS grants attackers direct access to the wireless physical layer, allowing to manipulate radio signals using simple software control. There is no other technology that offers on-the-fly wireless signal manipulation at comparable hardware cost and complexity (our prototype device has cost around USD~100).
The main contributions of this paper are listed as follows.
\begin{itemize}
\item  We propose a new kind of attack to CRKG, RIS jamming, and show that an attacker can prevent legitimate users from agreeing on the same key by deploying one or multiple malicious RISs to break channel reciprocity. In particular, we introduce three examples to realize the RIS-jamming attack, including using active nonreciprocal circuits, performing out-of-sync controls, and reducing signal-to-noise ratio (SNR). 
Unlike traditional attack methods that incur overhead by sending jamming signals, our proposed RIS-jamming attack takes advantage of the low cost and the passive nature of the RIS, which eliminates the need for sending signals, effectively avoiding the overhead problem.
  \item  We analyze the attack effect of RIS jamming by formulating the secret key rate with a relationship of the deployment of RIS. To resist such RIS-jamming attacks, we propose a countermeasure that exploits the wideband signals for multipath separation and then distinguishes the malicious path from all separated channel paths. Consequently, even in such difficult circumstances, it is still possible for legitimate users to generate secret keys by taking care of channel gains from the remaining paths.
  \item We verify the effectiveness of RIS-jamming attacks and their countermeasure through both simulations and experiments. Numerical simulation and experimental results are consistent, and they substantiate the statement that RIS-jamming attacks have a non-negligible negative effect on BDR and the proposed CPR-CRKG scheme can mitigate their impacts in wideband systems.
\end{itemize}

\section{Related Works}
\subsection{The Effects of RIS on CRKG}
A RIS is capable of creating an intelligent reconfigurable propagation environment. Thus, besides improving wireless communication,  
the RIS technology likewise has significant potential in view of CRKG~\cite{9771319,Gaoning_arXiv}. When the RIS is controlled by legitimate users, it can introduce an additional reciprocal channel path and enhance the temporal fluctuation of channel measurements. As a result, over the past three years, RIS has been considered a helper to assist CRKG in boosting the secret key generation rate.
With an elaborate design of RIS configuration, the secret key generation rate 
can be largely improved~\cite{21JiTVT,9771815,2021SPL,10197345}. For example, Ji~\textit{et al.} randomly changes the phase of RIS to introduce artificial randomness and increase the key generation rate (KGR)~\cite{21JiWCL}. 
In~\cite{9771815}, the authors optimized the RIS reflecting coefficients to maximize the key generation rate and minimize transmit power while guaranteeing the key generation rate target. 
RIS also makes it possible to yield secret keys from wireless channels in some harsh propagation scenarios, e.g., wave-blockage environments~\cite{9663196} and static or low-entropy environments~\cite{2020Intelligent}, extending the application range of PKG. 
The first field studies of RIS-assisted CRKG with an OFDM system were demonstrated by Staat~\textit{et al.} in~\cite{2020Intelligent}. Their prototype system achieved a key generation rate~(KGR) of 97.39~bps with a 6.5\% bit disagreement ratio~(BDR) after quantization while passing standard randomness tests in a static environment.

However, previous works rarely considered that RIS could likewise be controlled by an attacker. In the context of disrupting wireless communications, Staat~\textit{et al.} demonstrated how a RIS can disrupt Wi-Fi networks~\cite{StaatMirror2022}. In the context of CRKG, we have partially investigated an example of the RIS-jamming attack in our previous work~\cite{vtc_hu}. However, its attack model is not general and no field verification has been provided. It is yet to be explored, how this attack impacts CRKG in practical scenarios and how to withstand this attack.  

\subsection{Existing Jamming Attacks and Anti-Jamming Methods for CRKG}
Since CRKG is done over an open channel, it is an easy target for an adversary to focus all its jamming energy on the key establishment protocol. 
For this reason, jamming and anti-jamming techniques on CRKG have also been studied in the literature~\cite{08SPjamming,18TVTXiao,20TCOM_Ali,2020WN,17TIFSArsa,Zafer12ton,ESORICS12,zengkai}. 
Zeng~\emph{et~al.}~\cite{zengkai} gave a nice overview of the jamming of CRKG and other active attacks, including disruptive jamming, manipulative jamming, and channel manipulation (movement-based). 

Frequency Hopping Spread Spectrum (FHSS) is one of the most efficient mechanisms used for anti-jamming communication. For example, Xiao~\emph{et~al.}~\cite{18TVTXiao} proposed a fast deep Q-network (DQN) based two-dimensional (2-D) anti-jamming mobile communication system that applies both frequency hopping and user mobility to address both jamming and strong interference. However, as FHSS requires a pre-shared secret, the anti-jamming/key establishment becomes a circular dependency problem. 
In~\cite{08SPjamming}, Strasser \textit{et~al.} propose a Uncoordinated Frequency Hopping (UFH)-based mechanism to break this circular dependency, however, at the high cost of reducing the achievable rates for secret key establishment. To address this issue, \cite{17TIFSArsa} studied energy harvesting and FHSS technique over multiple orthogonal subcarriers to defend against jamming in CRKG procedure from the perspective of game theory. Further, the authors in~\cite{20TCOM_Ali,2020WN} considered a cooperative relaying scenario and proposed jamming-resistant schemes under the presence of an untrusted relay and an adversary jammer. They used the local observations of the two-hop channels to generate a shared secret key, which specified the adopted frequency hopping spread spectrum sequence to mitigate the effect of jamming on both the training and data transmission phases~\cite{20TCOM_Ali}. 
In addition to FHSS, Xie~\emph{et~al.}~\cite{23Xie}, 
cleverly exploited the difference of noise variances to defend against jamming attacks for the Physical-Layer Authentication (PLA)~\cite{9279294}.

Different from the above active jamming attacks which use their own internal energy to transmit strong noise signals to obstruct the key establishment and thus can be detected through the identification of increased energy levels, e.g. by detecting energy footprints, RIS-jamming attacks can use directly the signals of the victim system to attack by changing their reflection coefficients and phase shifts~\cite{RIS_J2020WCL}. 
Consequently, RIS-jamming attacks are very
difficult to detect and prevent since the distribution characteristics of RIS-jamming signals and noise differ significantly and discerning between RIS-jamming signals and those naturally reflected by the environment becomes challenging when relying solely on distinguishing the noise variances as used in~\cite{23Xie}. 
Actually, a RIS-jamming attack is even tougher as \textit{RIS jams like a shadow} -- whatever frequency Alice and Bob hop to, the jamming signal will follow. For this reason, the RIS-jamming attack can hardly be resolved by previous frequency hopping-based anti-jamming approaches.

To bridge these gaps, this paper carries out a comprehensive study of the RIS-jamming attack in the field of CRKG from both theoretical and experimental aspects.  
 
\section{System Model} 
\label{sec:model}
In this section, we build a general model for the scenario where a RIS participates in the CRKG process. On this basis, we give the problem statement.

\subsection{RIS-involved Channel Model}
\begin{figure}[h]
    \centering
    \includegraphics[width=0.9\linewidth]{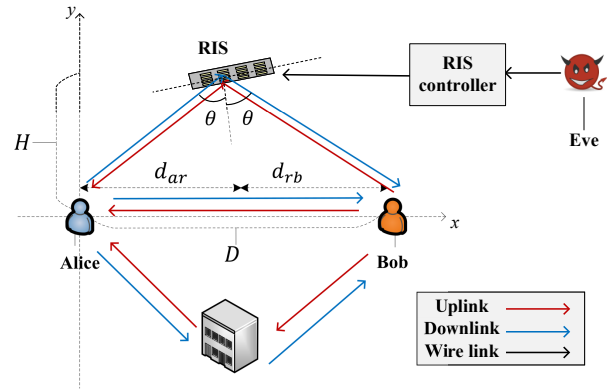}
    \caption{System model of the RIS-involved CRKG scenario.}
    \label{fig:system_model}
\end{figure}

Fig.~\ref{fig:system_model} depicts the RIS-involved CRKG scenario considered in this paper, considering three parties. 
Alice and Bob are two legitimate terminals that seek to establish a shared cryptographic key from their channel observations. For simplicity, they are assumed to be equipped with a single antenna and deploy a standard TDD wireless communication protocol, e.g., IEEE 802.11n Wi-Fi with orthogonal frequency-division multiplexing (OFDM) for bi-directional communications. 
Eve is an attacker who is aware of the key generation procedure of Alice and Bob. 
The distance between Eve and Alice/Bob is several orders of magnitude larger than the wavelength, and thus Eve's channel observations are assumed to be independent of those of Alice and Bob. 
In order to accomplish the purpose of interrupting the CRKG procedure, Eve places a malicious RIS device around to partially control the wireless propagation channel. The malicious RIS device is likely to be placed near Alice/Bob to guarantee a significant attack effect, which will be further discussed in Sec.~V.C.

To illustrate this effect, we consider a simplified 2D~system setup, as shown in Fig.~\ref{fig:system_model}, where Alice and Bob are located on a line with horizontal distance~$D$~m, and the RIS device is deployed on a line $H$~m above Alice and Bob. The horizontal distance between the RIS and Alice is denoted by $d_{ar}$. Accordingly, $d_{rb}=D-d_{ar}$ represents the horizontal distance between the RIS and Bob. The RIS consists of $M=M_x \times M_y$ reflecting elements of size $d_x \times d_y$. Each reflecting element is an electrically small low-gain element embedded on a substrate. 

In general, the total channel between Alice and Bob is the superposition of the direct link and the RIS-induced link. Accordingly, we elaborate on the model of these two links below. 
First, regarding the direct link, we suppose that it has $L$ propagation paths,
and its channel impulse response can be expressed as
\begin{align}
   g_d(\tau) = \sum_{\ell=1}^L \alpha_\ell \delta(\tau-\tau_\ell),
\end{align}
where $\alpha_\ell \in \Cb$ and $\tau_\ell \in \Rb$ are the complex channel gain and the time delay of the $\ell$-th path, respectively. 
The statistical power gain of each path, $\beta_\ell$, is decided by the path loss, which can be modeled as
\begin{align}
    \beta_\ell = \Eb\{ \alpha_\ell \alpha_\ell^* \} = 1/ d_\ell^2,
\end{align}
where $d_\ell$ is the distance of the $\ell$-th path.
Hence, its channel frequency response on the $k$-th subcarrier can be expressed as
\begin{align}
    h_d(k) = \sum_{\ell=1}^L \alpha_\ell  e^{-j2\pi \tau_\ell k/K }, k \in \{1,2, \cdots, K\},
\end{align}
where $K$ is the number of subcarriers.

Second, the channel impulse response of the RIS-induced channel is modeled as
\begin{align}\label{eq:4}
 h_{r}(\tau) = {\bf h}_{ar}^T {\bf \Lambda} {\bf h}_{rb}\delta(\tau-\tau_r),
\end{align}
where $\tau_r$ is the time delay of the RIS path,
${\bf h}_{ar} \in \Cb^{M\times 1}$ and ${\bf h}_{rb} \in \Cb^{M\times 1} $denote the channel coefficient vector from Alice to the RIS and the channel coefficient vector from the RIS to Bob, ${\bf \Lambda} = \Diag (\lambda_1, \lambda_2, \cdots, \lambda_M)$ is a diagonal matrix, representing the signal reflection of RIS. The elements of ${\bf \Lambda}$ are the equivalent reflection coefficients of each unit cell. 
Its $i$-th diagonal element is modeled as follows:
\begin{align}
\lambda_i = a_i e^{j\Phi_i}, i = 1, 2, \cdots, M.
\end{align}
where $a_i$ and $\Phi_i$ correspond to the amplitude response and the phase response, respectively.
Here, $a_i$ and $\Phi_i$ can be adjusted by a smart controller, e.g., a microcontroller or a field-programmable gate array (FPGA). 
In (\ref{eq:4}), the power gain of the RIS path is given by
\begin{align}\label{eq:6}
    \beta_r = \Eb\{({\bf h}_{ar}^T {\bf \Lambda}{\bf h}_{rb})({\bf h}_{ar}^T {\bf \Lambda} {\bf h}_{rb})^* \} = M \beta |\Gamma cos(\theta)|^2 / d_r^2,
\end{align}
where $\beta=\Eb\{ a_ia_i^*\}$ indicates the power of RIS units, $d_r$ is the distance of RIS path, $\theta$ is the incident angle. $\Gamma$ is the Fresnel reflection coefficient,
which can be modeled as~\cite{Wireless}
\begin{align}\label{eq:Gam}
    \Gamma = \frac{\epsilon_r \cos\theta - \sqrt{\epsilon_r - \sin^2\theta}}
    {\epsilon_r \cos\theta + \sqrt{\epsilon_r - \sin^2\theta}},
\end{align}
where $\epsilon_r$ denotes the relative permittivity of the reflecting medium. 
For simplicity, we assume that all reflection units have the same $\beta$ and $\Gamma$. By adding a power amplifier, $\beta$ can be magnified.

To sum up, the total channel impulse response can be expressed as
\begin{align}
    h(\tau) = \sum_{\ell=1}^L \beta_\ell \delta(\tau-\tau_\ell) 
    + \h_{ar}^T {\bf \Lambda}\h_{rb} \delta(\tau-\tau_r) 
\end{align}
and channel frequency response between Alice and Bob is given by 
\begin{align}\label{eq:7}
h(k) = h_d(k) + {\bf h}_{ar}^T {\bf \Lambda} {\bf h}_{rb} e^{-j2\pi \tau_r k /K}.
\end{align}
By stacking these channel frequency responses of all subcarriers together, we obtain the entire channel vector of 
\begin{align}\label{eq:8}
{\bf h} =[h(1), h(2), \cdots, h(K)]^T.
\end{align}

\subsection{Process of CRKG}
In the process of CRKG, Alice and Bob first exchange known OFDM symbols as probe signals to gather channel parameters. Each party can then use the received signal along with the probe signal to compute an estimate $\hat{\bf h}$ of $\bf h$. 
After channel probing, these channel estimates are then translated into identical bit-strings suitable for use as cryptographic keys. Since the channel estimates are continuous random variables, Alice and Bob first quantize them into raw keys using single-bit CDF quantization~\cite{2018trans}.
Next, the information reconciliation and privacy amplification procedures are used to generate identical and private secret keys. 

During the channel probing phase, for simplicity, we assume that all the above physical channels are block-fading. For the $k$-th subcarrier, its channel frequency response at time $t$ is given by 
\begin{align} 
h(k,t) = \sum_{n=-\infty}^\infty h_n(k)x(t-nT_c), 
\end{align}
where $h_n(k)$ is the channel frequency response for the $n$-th block and $x(t)$ is a rectangular wave
\begin{align}
x(t) = \begin{cases}
1, & nT_c\le t< (n+1) T_c, \\
0, & otherwise,
\end{cases}
\end{align}
and $T_c$ is the channel coherence time.

In a TDD system, Alice waits to receive a probe signal from Bob before responding with a probe signal and vice-versa. Assume that Alice and Bob transmit pilot signals at $t_1$ and $t_2$, respectively, the received signal at Bob and Alice at the $k$-th subcarrier can be written as
\begin{align}\label{eq:9}
   y_{ab}(k,t_1)& = h(k,t_1) s(t_1) + n_b(k,t_1) ,\\ \nonumber
   y_{ba}(k,t_2)& = h(k,t_2) s(t_2) + n_a(k,t_2),
\end{align} 
where $s(t_1)$ and $s(t_2)$ denote the known probe signal, $n_b(k,t_1)$ and $n_a(k,t_2)$ are the independent additive white Gaussian noise (AWGN) processes at Bob and Alice. In this paper, the SNR is defined as the ratio of pilot signal power to noise power.

Using the received signal, Bob and Alice utilize a least-squares (LS) channel estimation: 

\begin{align} \label{eq:15}
    \tilde h_{ab}(k,t_1)=&{\bf h}_{ar}^T(t_1) {\bf \Lambda} (t_1){\bf h}_{rb}(t_1)e^{-j2\pi \tau_r(t_1) k /K} \nn\\
    &+ h_{d}(k,t_1) + z_b(k,t_1) , \nonumber \\
    \tilde h_{ba}(k,t_2)=&{\bf h}_{rb}^T(t_2) {\bf \Lambda}(t_2){\bf h}_{ar}(t_2)e^{-j2\pi \tau_r(t_2) k /K}  \nn\\
    &+h_{d}(k,t_2) + z_a(k,t_2),
\end{align} 
where ${\bf \Lambda} (t_2)$ is the RIS coefficients at time $t_2$, which is independent from 
${\bf \Lambda} (t_1)$. The statistical gain of the RIS path at $t_2$ is $\beta_{r'}$. $z_b(k,t_1)$ and $z_a(k,t_2)$ in (\ref{eq:15}) represent the AWGN terms due to $n_b(k,t_1)$ and $n_a(k,t_2)$ after processing by the function that estimates $h$. They both have a variance of $\sigma^2$ so that the SNR is equal to $1/\sigma^2$.

In one channel probing round, the time difference $t_2-t_1$ is smaller than 
$T_c$, guaranteeing a high correlation of the channel observations. 

The last three steps after channel probing are similar to existing works in the field of CRKG and are thus not particularly designed in this paper. We refer the interested reader to~\cite{zenger2015line,zenger2015preventing,zenger2016preventing,zenger2017physical}.

\subsection{Problem Statement}
According to the above system model, the entire channel is partially under the control of Eve. Previous research indicates that the artificial electromagnetic characteristics of RISs do not strictly follow the normal laws of reciprocity~\cite{Tang2021}. As a result, Eve is able to obstruct the secret key agreement of CRKG by reducing the similarity of the legitimate bidirectional channel estimates. 
Although this attack idea is initially formed, more problems need to be investigated: 
\begin{itemize}
\item How to design or control a RIS to realize this kind of attack in practice?

\item How much of an impact can Eve have on the CRKG between Alice and Bob? 

\item Is it possible for Alice and Bob to mitigate the potential severity of this attack?
\end{itemize}
The rest of this paper will seek solutions to these questions from both theoretical and experimental aspects.

\section{RIS-jamming attack} 
\label{sec:ris_attack_scheme}
In this section, we first describe the basic idea of the RIS-jamming attack and then elaborate on three implementation examples.

\begin{figure*}
\subfigure[The asymmetry of RIS structures.]{
\begin{minipage}[c][5cm][c]{.26\textwidth}
\centering
\includegraphics[width=1.0\linewidth]{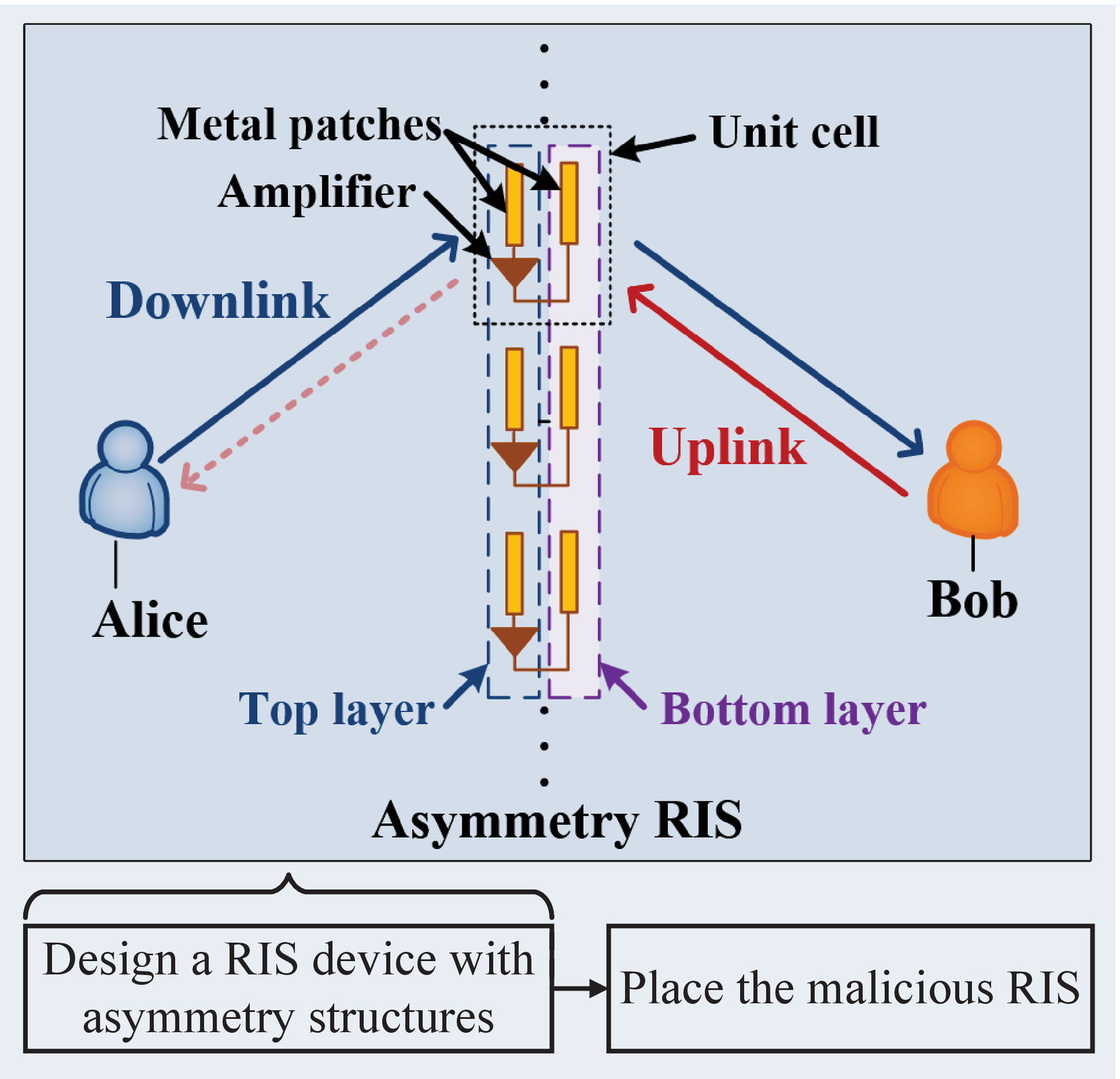}
\end{minipage}
}
\subfigure[Asynchronous RIS configurations.]{
\begin{minipage}[c][5cm][c]{.33\textwidth}
\centering
\includegraphics[width=1.0\linewidth]{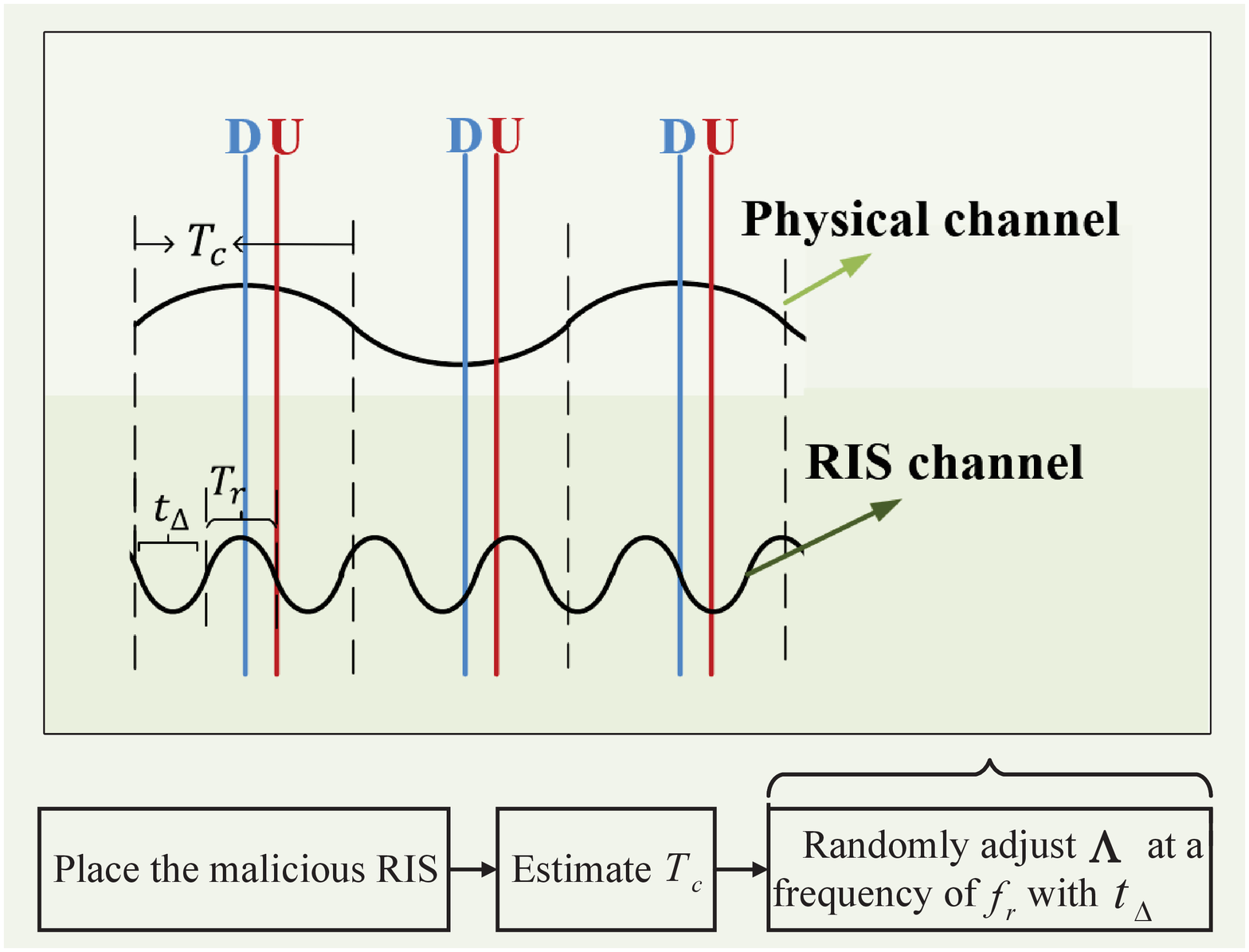}
\end{minipage}
}
\subfigure[RIS-enhanced direct link cancellation design.]{
\begin{minipage}[c][5cm][c]{.31\textwidth}
\centering
\includegraphics[width=1.1\linewidth]{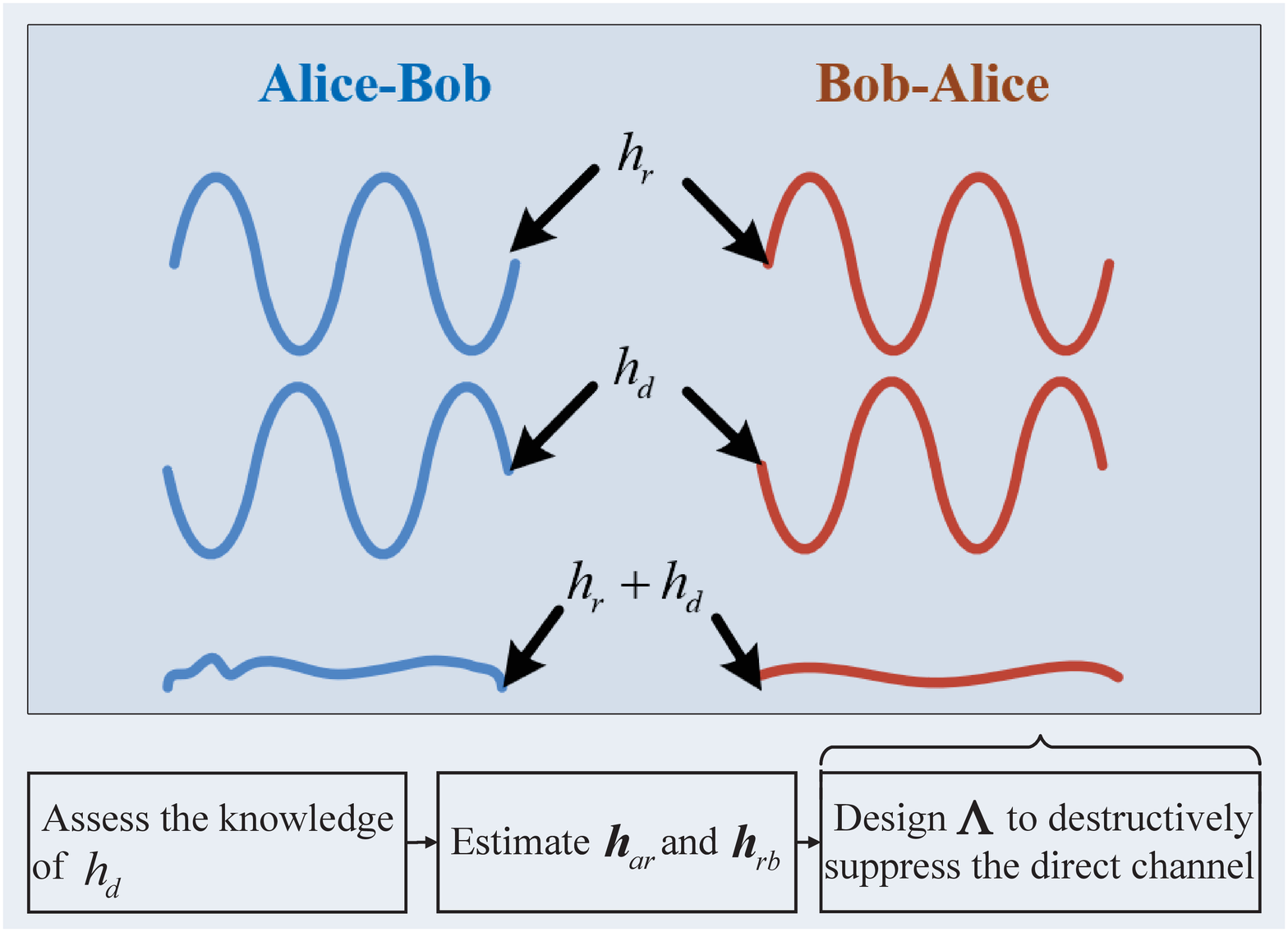}
\end{minipage}
}
\caption{Illustrations of the three RIS-jamming attacks.}
\label{Fig.2}
\end{figure*}

\subsection{Adversary Model}
In this paper, we focus on an adversary, Eve, whose goal is to jam and break the reciprocity of the communication between the legitimate parties, Alice and Bob. Instead of eavesdropping the secret information in passive attacks, Eve strategically employs a RIS to execute active attacks to interfere with the seamless exchange of information. It is essential to note that the malicious RIS should be placed near Alice/Bob to guarantee a high reflectivity and a small signal attenuation, ultimately leading to the breakdown of the key generation process. We consider three types of RIS-jamming attacks, including using active nonreciprocal circuits, performing asynchronous controls, and reducing the SNR. Further details will be explored in subsequent subsections. 

\subsection{Idea Description} 
Regarding the system model as described in Fig.~\ref{fig:system_model}, the success of CRKG relies on the reciprocity of $ \tilde h_{ab}(t_1, k)$ and $\tilde h_{ba}(t_2, k)$. When their reciprocity is weak, the BDR of the raw key becomes high.
%
%
{In this situation, correcting the errors via information reconciliation results in a great burden of the high computation cost and significant information leakage of the secret key.}
Typically, it is difficult to obtain identical key material with CRKG if the BDR of the raw key is larger than 20\%. Building on this observation, a RIS-jamming attack may obstruct the CRKG between Alice and Bob in two ways. 
\begin{itemize}
\item {\textit {Breaking the reciprocity of the RIS-induced link:}} 
As observed from (\ref{eq:15}), the reciprocity of the channel estimates depends on the reciprocity of the direct link as well as the RIS-induced link. 
If Alice and Bob send probes to one another at a fast enough rate, i.e., if $T_p = t_2 - t_1$ is small, the physical channels between Alice and Bob at $t_1$ and $t_2$ can be highly correlated. The time delays 
are assumed to be the same. 
However, this high correlation is not guaranteed for ${\bf \Lambda} (t_1)$ and ${\bf \Lambda} (t_2)$ as these RIS-related parameters are under the control of Eve. In other words, Eve can reduce the correlation of ${\bf \Lambda} (t_1)$ and ${\bf \Lambda} (t_2)$ by applying intentional changes of the RIS response at time instants $t_1$ and $t_2$. 
\item {\textit {Reducing the SNR of the entire channel:}} The reciprocity of the entire channel estimates is also affected by the SNR due to the independent noise terms $z_a(k,t_2)$ and $z_b(k,t_1)$ in (\ref{eq:15}). Thus, another approach for Eve to break key agreements is to build a RIS-induced link that can suppress the direct link. In this scenario, the reciprocal channel is submerged by noise and thus makes it difficult to reach a key agreement. 
In the worst case, the RIS-induced channel cancels the direct channel, leaving only independent noise in the channel estimates. 
\end{itemize}
In both cases, the reciprocity of $ \tilde \h_{ab}(t_1)$ and $\tilde \h_{ba}(t_2)$ will be disrupted due to the RIS operation. {For this reason, we define such kind of attack as a RIS-jamming attack in this paper.} 

\subsection{Attack Model}
\label{sec:three}
Considering there are only three parameters of RIS that can be modified, i.e., the hardware structure, the switching frequency, and the reflection coefficients of the RIS, the attacker can only implement a RIS-jamming attack through three aspects. Thus, we elaborate on three corresponding examples of the RIS-jamming attack. Fig.~\ref{Fig.2} illustrates their principles and processing flows. The first two of them aim at breaking the reciprocity of the RIS-induced link, while the last one intends to reduce the SNR of the entire channel. 
\subsubsection{Asymmetry of RIS Structures}
The first way of implementing a RIS-jamming attack makes use of the asymmetry of the RIS structure to cause different RIS signal reflection matrices in forward and reverse channels, while the reflection matrix coefficient of RIS is just randomly configured. That is, the response of a RIS unit cell can vary with the angle of incidence, thus yielding a non-reciprocal response~\cite{9206044}. Further, RIS unit cells could be deliberately designed to behave non-reciprocal, e.g., by utilizing active non-reciprocal circuits such as microwave amplifiers or isolators. 

For example, as depicted in Fig.~\ref{Fig.2}(a), in the forward direction of the RIS-induced link at time $t_1$, wireless signals are first captured by the metal patches on the RIS’s top layer and transformed into circuit signals. After passing through amplifiers, the circuit signals are then radiated into space again by the metal patches on RIS’s bottom layer.
In the reverse direction at time $t_2$, however, due to the non-reciprocity of integrated amplifiers, the signals captured by the metal patches on RIS’s bottom layer are isolated. In this case, the channels estimated at Bob and Alice are mathematically given by

\begin{align} \label{eq:13}
    \tilde h_{ab}(k,t_1)=&{\bf h}_{ar}^T(t_1) {\bf \Lambda} (t_1){\bf h}_{rb}(t_1)e^{-j2\pi \tau_r k /K} \nn\\
    &+ h_{d}(k,t_1) + z_a(k,t_1) , \nonumber \\
    \tilde h_{ba}(k,t_2)=&h_{d}(k,t_2) + z_b(k,t_2).
\end{align} 
As observed, $ \tilde h_{ba}(k,t_2)$ differs from $ \tilde h_{ba}(k,t_1)$, {mainly due to the isolation of its RIS-induced link,} i.e., ${\bf \Lambda} (t_2)=0$. 
\subsubsection{Asynchronous RIS Configurations}
The second method intends to make the channel observations of Alice and Bob non-reciprocal rather than introducing an actual non-reciprocal RIS link. For this, Eve takes advantage of asynchronous RIS configurations. 
Here, Eve randomly configures the RIS to adjust ${\bf \Lambda}$ at a frequency of $f_r$.
In other words, the reflection coefficients of the RIS are randomly configured and remain constant within a time block of $T_{r} = 1/f_r$.  
Then, the equivalent reflection coefficient of the $i$-th unit cell of the RIS at time $t$ is given by 
\begin{align}\label{eq:17}
\lambda_i (t) = \sum_{n=-\infty}^\infty \lambda_{i,n}x'(t-nT_{r}+t_{\Delta}), 
\end{align}
where $\lambda_{i,n}$ is the equivalent reflection coefficient for the $n$-th time block and $x'(t)$ is also a rectangular wave 
\begin{align}\label{eq:18}
x'(t) = \begin{cases}
1, & nT_{r}\le t< (n+1) T_{r}, \\
0, & otherwise,
\end{cases}
\end{align}
and $t_{\Delta}$ is the initial temporal deviation between $\lambda_i$ and the physical channel. 
When $T_r = T_c$ and $t_{\Delta} = 0$, the change in RIS is synchronized with the change of physical channels. In this case, ${\bf \Lambda} (t_1) = {\bf \Lambda} (t_2)$, which will not have negative effects on CRKG. Conversely, if $T_r < T_c$ or $t_{\Delta} \ne 0$, ${\bf \Lambda} (t_1)$ will be different from ${\bf \Lambda} (t_2)$ with high probability. 
Accordingly, as illustrated in Fig.~\ref{Fig.2}(b), Eve can realize the asynchronism by accelerating the switching frequency of the RIS, introducing a considerable temporal deviation, or both. In either case, $t_1$ and $t_2$ fall into different time block, so that ${\bf \Lambda} (t_1)$ will be independent from ${\bf \Lambda} (t_2)$, which reduces the similarity between $ \tilde h_{ba}(k,t_2)$ and $ \tilde h_{ba}(k,t_1)$.    
\subsubsection{RIS-enhanced Direct Link Cancellation Design}
Another example of the RIS-jamming attack in CRKG is direct link cancellation, where the reflected channel and the direct channel can be destructively combined, as shown in Fig.~\ref{Fig.2}(c). 
However, this example requires that Eve knows not only the channel from Alice/Bob to the RIS but also the direct link between Alice and Bob, which is a strong assumption in practice. One possible case is in a static environment and Eve performs the Position Replay Attack~\cite{Applying_12TIFS,9796694} in which Eve records the position of Bob and moves to it after Bob leaves to acquire similar channel measurements. Additionally, once Eve knows the direct link between Alice and Bob, after weighing the advantages and disadvantages, she would prefer to steal the secret key insensibly rather than destroy the key generation. 

\begin{table*}
	\label{Three ways of RIS-jamming attack}
	\centering
	\caption{The comparison of three RIS-jamming attacks.}
	\scalebox{0.98}{
	\begin{tabular}{| c | c | c | c | c | c |}
	\hline
			& RIS parameters that Eve can modify & Requirement & Destructiveness 
			& Implementation & Target
			\\
			\hline
			\multirow{2}*{1} & {Hardware structure} &\multirow{2}*{Active non-reciprocal circuits on RIS} & \multirow{2}*{Middle} & \multirow{2}*{Middle} & {Break the}
			\\ 
			&(e.g., asymmetric RIS structure) & & & &{reciprocity of}
			\\
			\cline{1-5}
			\multirow{2}*{2} & Switching frequency & \multirow{2}*{Fast switching speed of RIS configurations} & \multirow{2}*{Strong} & \multirow{2}*{Easy} & {RIS-induced}
			\\
			& (e.g., asynchronous RIS configuration)& & & & link
			\\
			\hline
			\multirow{2}*{3} & Reflection coefficients & \multirow{2}*{Knowing the direct link} & \multirow{2}*{Very strong} & \multirow{2}*{Extremely difficult}&  Reduce the SNR of
			\\
			& (e.g., direct link cancellation) & & & &the entire channel
			\\
			\hline 
	\end{tabular}}
\end{table*}    

Table~\ref{Three ways of RIS-jamming attack} summarizes the above three examples of the RIS-jamming attack from the aspects of technical means, requirement, destructiveness, and implementation. 
Considering the last example has strict requirements for implementation, the rest of this paper is focused on the first two examples.
\section{The Analysis of Attack Effects} \label{sec:analysis}
To evaluate the effect of the RIS-jamming attack on CRKG, we formulate the secret key rate and then discuss effective regions of RIS deployment.
\subsection{The Secret Key Rate} 
In general, the number of secure bits yielded by CRKG is the mutual information between $\tilde{h}_{ab}$ and $\tilde{h}_{ba}$ on the condition of $\tilde{h}_{e}$ that can be expressed as~ \cite{maurer1993secret}
\begin{align}
    R = I(\tilde{h}_{ab};\tilde{h}_{ba}|\tilde{h}_{ae};\tilde{h}_{be}),
\end{align}
where $\tilde{h}_{ae}$ and $\tilde{h}_{be}$ are the channels observed by Eve.
As $\tilde{h}_{ae}$ and $\tilde{h}_{be}$ are assumed to be independent of those of Alice and Bob, the secret key rate degrades to 
\begin{align}\label{eq:21}
    R = I(\tilde{h}_{ab};\tilde{h}_{ba})= \log \frac{K_{ab}K_{ba}}{\det(\K_{ab})},
\end{align}
where $K_{ab}$ and $K_{ba}$ are respectively covariances of $\tilde{h}_{ab}$ and $\tilde{h}_{ba}$, and $\K_{ab}$ is the total correlation matrix which is given by
\begin{align}
    \K_{ab} = \Eb \begin{bmatrix}
        \tilde{h}_{ab}\tilde{h}_{ab}^*  & \tilde{h}_{ab}\tilde{h}_{ba}^* \\
        \tilde{h}_{ba}\tilde{h}_{ab}^*  & \tilde{h}_{ba}\tilde{h}_{ba}^*
    \end{bmatrix}.
\end{align}
For convenience of calculations, we omit $t_1$ and $t_2$ and simplify channel estimates of (\ref{eq:15}) into  
\begin{align} \label{eq:24}
    \tilde h_{ab}(k)&={\bf h}_{ar}^T{\bf \Lambda} {\bf h}_{rb}e^{-j2\pi \tau_r k /K}+ h_{d}(k) + z_b(k) ,  \\ \nonumber
    \tilde h_{ba}(k)&={\bf h}_{rb}^T{\bf \Lambda}'{\bf h}_{ar}e^{-j2\pi \tau_r k /K}+h_{d}(k) + z_a(k).
\end{align} 
Then, $K_{ab}$ and $K_{ba}$ are calculated as
\begin{align}
    K_{ab} &= \Eb\{ \tilde{h}_{ab}\tilde{h}_{ab}^* \} \nn\\
    &= \Eb\{ |\h_{ar}^T\Lambdam\h_{rb}|^2 + h_d h_d^* + \sigma^2 \} \nn\\
    &= \beta_r + \sum_{l=1}^L \beta_\ell + \sigma^2,
\end{align}
and 
\begin{align}
    K_{ba} &= \Eb\{ \tilde{h}_{ba}\tilde{h}_{ba}^* \} \nn\\
    &= \Eb\{ |\h_{rb}^T\Lambdam' \h_{ar}|^2 + h_d h_d^* + \sigma^2  \} \nn\\
    &= \beta_{r'} + \sum_{l=1}^L \beta_\ell + \sigma^2,
\end{align}
respectively.  

The cross-terms satisfy $\Eb\{\tilde{h}_{ba}\tilde{h}_{ab}^*\}=\Eb\{\tilde{h}_{ab}\tilde{h}_{ba}^*\}=K_c$, which is calculated as
\begin{align}
K_c &=  \Eb\{{\bf h}_{ar}^T {\bf \Lambda} {\bf h}_{rb} ({\bf h}_{rb}^T {\bf \Lambda}'{\bf h}_{ar})^*    + h_d h_d^*\} \\ \nonumber
&= \beta_{r,r'}+ \sum_{l=1}^L \beta_\ell,
\end{align}
where $\beta_{r,r'}=\Eb\{{\bf h}_{ar}^T {\bf \Lambda} {\bf h}_{rb} ({\bf h}_{rb}^T {\bf \Lambda}'{\bf h}_{ar})^*\}$ represents the cross-covariance of the RIS-induced link. 
Then, the determinant of $\K_{ab}$ can be calculated as
\begin{align}
    \det(\K_{ab}) = &(\beta_r + \sum_{l=1}^L \beta_\ell + \sigma^2)
    (\beta_{r'} + \sum_{l=1}^L \beta_\ell + \sigma^2)  \nn\\
    &- (\sum_{l=1}^L \beta_\ell +\beta_{r,r'})^2 .
\end{align}
Hence, the secret key rate can be expressed as

\begin{align}\label{eq:29}
R=&\log\Big(\beta_r + \sum_{l=1}^L \beta_\ell + \sigma^2\Big)+\log\Big( \beta_r' + \sum_{l=1}^L \beta_\ell + \sigma^2 \Big)\nn\\
&- \log\Big((\beta_r + \sum_{l=1}^L \beta_\ell + \sigma^2)( \beta_r' + \sum_{l=1}^L \beta_\ell + \sigma^2) \nn\\
&\qquad ~~~ -(  \beta_{r,r'}+\sum_{l=1}^L \beta_\ell )^2\Big).
\end{align}
Notably, with the increase of path number $L$, the energy proportion of the RIS path declines and the attack effect diminishes. In other words, a RIS-jamming attack is more destructive in the scenario where the scattering environment is not rich enough.
\subsection{Case Studies}\label{sec:VB}
Next, according to (\ref{eq:29}), we derive secret key rates of some specific cases to illustrate the effect of the RIS on CRKG:
\begin{itemize}
\item {\textit {No RIS is deployed:}} In the absence of a RIS, $\beta_{r,r'}=\beta_{r'}=\beta_{r}=0$, and then the secret key rate is given by
\begin{align}\label{eq:30}
R_0= \log \Big( 1 + \frac{(\sum_{l=1}^L \beta_\ell)^2 }{\sigma^4 
    + 2 \sigma^2 (\sum_{l=1}^L \beta_\ell )} \Big),
\end{align}
which serves as a baseline for this paper. For the sake of analysis, we assume that $\sum_{l=1}^L \beta_\ell$ is fixed and discuss the change of secret key rates with $\beta_{r}$ and $\sigma^2$.
\item {\textit {RIS is not malicious:}} Next, we consider the case that a RIS exists, but \textbf{it is not controlled by Eve}. The reflection coefficients of RIS are assumed to be constant during one channel probing period, i.e., ${\bf \Lambda}' = {\bf \Lambda}$, so $\beta_{r,r'}=\beta_{r'}=\beta_{r}$, and the secret key rate becomes
\begin{align}\label{eq:31}
R_1=\log \Big( 1 + \frac{(\beta_{r}+\sum_{l=1}^L \beta_\ell )^2}{
 \sigma^4+ 2\sigma^2(\beta_r + \sum_{l=1}^L \beta_\ell)} \Big).
\end{align}
As the power gain of the RIS path, $\beta_{r}$, is positive and the basic function
\begin{align}
f(x)=\frac{x^2}{a+bx}, a>0, b>0, x>0
\end{align}
increases monotonously along with $x$, we obtain that $R_1 > R_0$, which indicates that instead of decreasing, a RIS will even increase the secret key rate as long as the change of the RIS path is synchronized with those of physical paths. 
\item {\textit {RIS-Jamming Attacks:}} Now, we study the secret key rate under RIS-jamming attacks. First, we consider that Eve exploits \textbf{the asymmetry of RIS structures}, i.e., ${\bf \Lambda}'= 0$, so $\beta_{r,r'}=\beta_{r'}=0$ and the secret key rate is derived as
\begin{align}\label{eq:32}
R_2 = \log \Big( 1 + \frac{( \sum_{l=1}^L \beta_\ell )^2}{\sigma^4+2 \sigma^2 \sum_{l=1}^L \beta_\ell +\beta_r \sum_{l=1}^L \beta_\ell + \beta_r \sigma^2} \Big).
\end{align}
Next, we consider another RIS-jamming attack, where Eve uses \textbf{asynchronous RIS configurations}. In this case, ${\bf \Lambda}'$ is independent of ${\bf \Lambda}$ and they have the same covariance, so that $\beta_{r,r'}=0$, $\beta_{r'}=\beta_{r}$ and the secret key rate becomes 
\begin{align}\label{eq:34}
R_3 = \log \Big( 1 + \frac{(\sum_{l=1}^L \beta_\ell )^2}{(\beta_r + \sigma^2)^2 +2(\beta_r + \sigma^2)( \sum_{l=1}^L \beta_\ell )} \Big).
\end{align}
\end{itemize}

Comparing (\ref{eq:32}) and (\ref{eq:30}), it is observed that $R_2$ is smaller than $R_0$, since the additional term $\beta_r \sum_{l=1}^L \beta_\ell + \beta_r \sigma^2$ in the denominator of (\ref{eq:32}) is positive. When the SNR is high, the gap between $R_0$ and $R_2$ is approximately 

\begin{align}\label{eq:33}
\Delta_1 =& R_0 - R_2 \nn\\
 \approx & \log\Big(\frac{(\sum_{l=1}^L \beta_\ell)^2}{\sigma^4+2\sigma^2 \sum_{l=1}^L \beta_\ell}\Big) \nn\\
&-\log\Big(1+\frac{\sum_{l=1}^L \beta_\ell}{\beta_r}\Big)\nn\\
&\approx \log\Big(\frac{\sum_{l=1}^L \beta_\ell}{2\sigma^2 }\frac{\beta_r}{\beta_r+\sum_{l=1}^L \beta_\ell}\Big).
\end{align}
Therefore, this gap can be enlarged by increasing $\beta_r$ or by decreasing $\sigma^2$. When the SNR is smaller than 0 dB, both $R_0$ and $R_2$ tend to be zero and the gap between them also approaches zero. For the intermediate SNRs, it is difficult to achieve an accurate approximation of $\Delta_1$, so we exploit (\ref{eq:33}) to analyze the trends of $\Delta_1$ versus SNR and $\beta$ approximately.
 
Similarly, we find that $R_3$ is smaller than $R_0$, since the additional term $2\beta_r\sum_{l=1}^L \beta_\ell+2\beta_r\sigma^2+ \beta_r^2$ in the denominator of (\ref{eq:34}) is positive. In addition, $R_3$ is also smaller than $R_2$, as $2\beta_r\sum_{l=1}^L \beta_\ell+2\beta_r\sigma^2+ \beta_r^2>\beta_r \sum_{l=1}^L \beta_\ell + \beta_r \sigma^2$, which indicates that as a RIS-jamming attack, asynchronous RIS configurations works better than asymmetric RIS structures. 

Define $\Delta_2 =R_0-R_3$ as the gap between this case and the baseline, and let $\Delta_3=R_2-R_3$ denote the gap between the two RIS-jamming attacks. 
At high SNR, these values are approximated by 
\begin{align}\label{eq:35}
\Delta_2  \approx& \log\Big(\frac{(\sum_{l=1}^L \beta_\ell)^2}{\sigma^4+2\sigma^2 (\sum_{l=1}^L \beta_\ell)}\Big)\nn\\
&-\log\Big(1+\frac{(\sum_{l=1}^L \beta_\ell)^2}{\beta_r^2+2\beta_r(\sum_{l=1}^L \beta_\ell)}\Big)\nn\\
\approx& \log\Big(\frac{(\sum_{l=1}^L \beta_\ell)\beta_r}{2\sigma^2}\frac{\beta_r+2(\sum_{l=1}^L \beta_\ell)}{(\beta_r+\sum_{l=1}^L \beta_\ell)^2}\Big)\nn\\
=&\log\Big(\frac{\sum_{l=1}^L \beta_\ell(1-\frac{1}{\rho^2})}{2\sigma^2}\Big),
\end{align}
and 
\begin{align}\label{eq:36}
\Delta_3 & \approx \log \Big(1+\frac{\sum_{l=1}^L \beta_\ell}{\beta_r}\Big)
-\log\Big(1+\frac{(\sum_{l=1}^L \beta_\ell)^2}{\beta_r^2+2\beta_r(\sum_{l=1}^L \beta_\ell)}\Big)\nn\\
&=\log\Big(1+\frac{\sum_{l=1}^L \beta_\ell}{\beta_r+\sum_{l=1}^L \beta_\ell}\Big)\nn\\
&=\log\Big(1+\frac{1}{1+\frac{\beta_r}{\sum_{l=1}^L \beta_\ell}}\Big)\nn\\
&=\log\Big(1+\frac{1}{\rho}\Big),
\end{align}
respectively, where $\rho=1+\frac{\beta_r}{\sum_{l=1}^L \beta_\ell}$.
As observed, $\Delta_2 $ increases with the decline of $\sigma^2$. It also rises with the increase of $\beta_r$ and it tends to be constant in the end.
Conversely, $\Delta_3 $ only decreases with the power of the RIS path
$\beta_r$. 

\begin{figure}[!h]
    \centering  \includegraphics[width=0.8\linewidth]{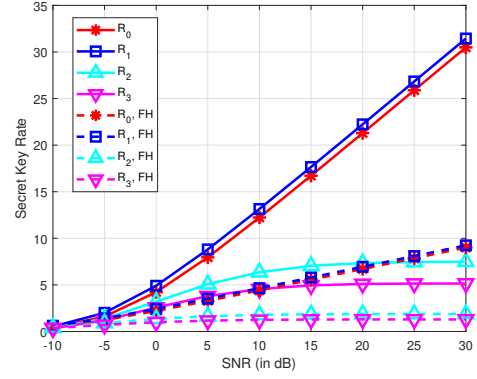}
    \caption{Different secret key rates versus SNR ($\beta = 1$).}
    \label{fig:Rate_SNR}
\end{figure}

\begin{figure}[!h]
    \centering
\includegraphics[width=0.8\linewidth]{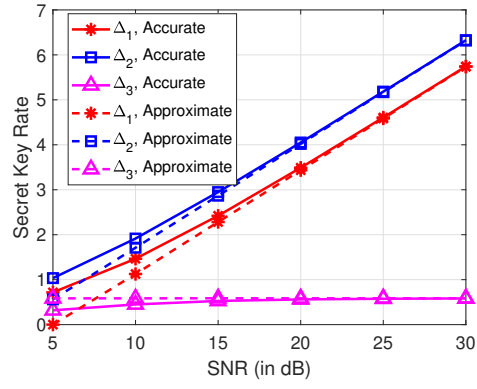}
    \caption{The differences between different secret key rates versus different SNR ($\beta = 1$).}
    \label{fig:Delta_SNR}
\end{figure}

\begin{figure}[!h]
    \centering
    \includegraphics[width=0.8\linewidth]{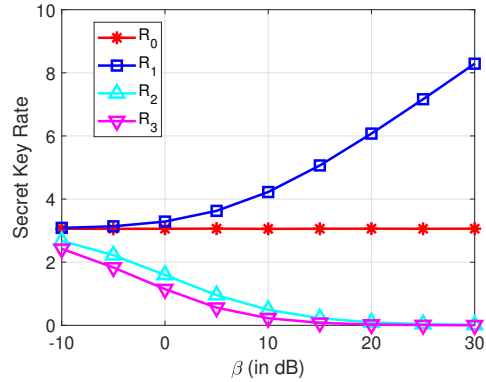}
    \caption{Different secret key rates versus $\beta$ (SNR = 10 dB).}
    \label{fig:Rate_beta}
\end{figure}

\begin{figure}[!h]
    \centering
    \includegraphics[width=0.8\linewidth]{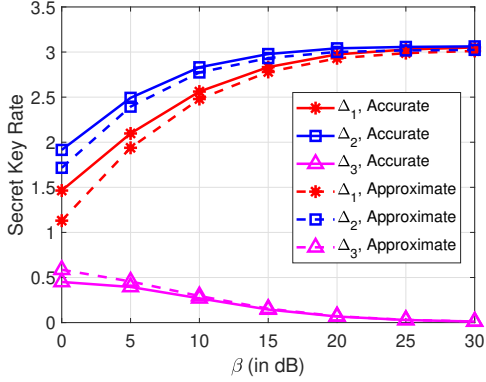}
    \caption{The differences between different secret key rates versus $\beta$ (SNR = 10 dB).}
    \label{fig:Delta_beta}
\end{figure}
Next, we use Monte Carlo simulations~\cite{tranter2004principles} to present numerical results of secret key rates to verify the above theoretical analysis. 
Without loss of generality, we consider a close-range CRKG scenario and assume that the number of paths is $L=4$. The distance of each path is generated according to the uniform distribution of the scatter in a range of 10~m~$\times$~10~m. The number of RIS elements is set 10. Following~\cite{sr}, in the simulation, the dielectric substrate of the meta-atom in the RIS is assumed to be RO4003 with a relative permittivity of 3.55. 
Figure~\ref{fig:Rate_SNR} reports the secret key rates, including $R_0$, $R_1$, $R_2$, and $R_3$, as functions of the SNR, which is equal to $1/\sigma^2$. The RIS power is set as $\beta = 1$. It is observed that $R_0$ and $R_1$ are significantly higher than $R_2$ and $R_3$, which is in agreement with the previous theoretical analysis that RIS-jamming attacks can reduce the secret key rate. Although all curves grow with SNR, the curves of $R_0$ and $R_1$ have almost linear increases, while the curves of $R_2$ and $R_3$ become less steep when the SNR is larger than 20 dB.
Specifically, $R_1$ is slightly larger than $R_0$ as a harmless RIS would provide an additional channel path for key generation. Regarding RIS-jamming attacks, $R_3$ is one-fourth lower than $R_2$ since asynchronous RIS configuration would affect both forward and backward RIS-induced channels while the asymmetry of RIS structure only has an impact on one direction. 
In addition, the secret key rates with FH are lower than those without FH, which is caused by the fact that only one subcarrier is used each time for the FH approach. Still, the FH approach has a good ability to overcome the noise as all energy is concentrated on one subcarrier. For this reason, the rate gap narrows with the decrease of SNR.
Figure~\ref{fig:Delta_SNR} further demonstrates the differences between these secret key rates, including accurate values and approximate values as given in (\ref{eq:33}), (\ref{eq:35}) and (\ref{eq:36}).
These approximate values are proved to be good as they are very close to accurate values at high SNR. 
Even for the SNR between 5 dB to 15 dB, these approximate values also demonstrate similar trends versus SNR as those of the true values.
$\Delta_1$ and $\Delta_2$, representing the rate decrease caused by the above two RIS-jamming attacks, rise almost linearly with SNR, while their difference, i.e., $\Delta_3$, is basically fixed over SNR. These curves are consistent with the theoretical analysis of $\Delta_1$, $\Delta_2$, and $\Delta_3$. 

In Fig.~\ref{fig:Rate_beta}, we compare the secret key rates for SNR = 10 dB as a function of the RIS units power $\beta$, which can be enlarged by adding a power amplifier. The power of RIS path $\beta_r$ increases with $\beta$ linearly. 
The original secret key rate $R_0$ without RIS is not affected by $\beta$, while $R_1$ rises with $\beta$ due to the power increase of the additional channel path provided by a harmless RIS. Conversely, if RIS is malicious, the secret key rates $R_2$ and $R_3$ drop with the growth of $\beta$ rapidly, approaching close to zero when $\beta$ surpasses 10 dB. The curve of $R_3$ falls even faster than that of $R_2$. The result in Fig.~\ref{fig:Rate_beta} substantiates the statement that the negative effect of RIS-jamming attacks on the secret key rate can be enhanced by increasing~$\beta$. 
Figure~\ref{fig:Delta_beta} further compares the differences of secret key rates as a function of $\beta$. The approximate values of $\Delta_1$, $\Delta_2$, and $\Delta_3$ are close to their corresponding accurate values, especially when $\beta$ is larger than 10 dB. The curves of $\Delta_1$ and $\Delta_2$ increase with $\beta$, while the curve of $\Delta_3$ decreases with $\beta$. 
This is caused by the fact that when $\beta$ grows, the effects of RIS-jamming attacks become more aggressive and the secret key rates, including $R_2$ and $R_3$, drop closer to zero.

From the above discussion, we have found that RIS-jamming attacks can reduce the secret key rate and their negative effects will be enhanced by increasing the power gain of the RIS path. 
\subsection{The Effective Region for RIS Deployment} 
\label{subsec:secret_key_rate_under_ris_attack}
Recall that the power gain of the RIS path $\beta_r = \beta |\Gamma cos(\theta)|^2/d_r^2$, in addition to $\beta$, $\beta_r$ also varies with position. Therefore, from Eve's point of view, she should choose an appropriate position of RIS to enhance its attack effects.

According to (\ref{eq:6}) and (\ref{eq:Gam}), the channel gain of the RIS path is
\begin{align}\label{eq:38}
    \beta_r = \beta\frac{\cos^2\theta}{d_r^2} \big| \frac{\epsilon_r \cos\theta - \sqrt{\epsilon_r - \sin^2\theta}}
    {\epsilon_r \cos\theta + \sqrt{\epsilon_r - \sin^2\theta} } \big|^2, 
\end{align}
which depends on its propagation distance $d_r$ and incident angle $\theta$. The further the distance, the greater the attenuation. 
Generally, the bigger the angle, the smaller the reflectivity. Therefore, if Eve would like to increase $\beta_r$, the RIS should be placed at a position {where the RIS path has a small propagation distance and a small incident angle.}

To find more insights on RIS positions, we further express $\beta_r$ as a function of $d_{ar}$ and $H$. 
As shown in Fig.~\ref{fig:system_model},  $d_r$ is the sum of the distance between Alice and RIS and that between RIS and Bob, i.e., 
\begin{align}\label{eq:39}
    d_r = \sqrt{d_{ar}^2 + H^2} + \sqrt{(D-d_{ar})^2 + H^2}.
\end{align}
The double angle $2\theta$ satisfies the Cosine law that
\begin{align} \label{eq:40}
    \cos2\theta &=\frac{\left ( {d_{ar}}^2+H^2 \right )+\left [ \left ( D-d_{ar} \right )^2+H^2  \right ] -D^2 }{2\cdot\sqrt{{d_{ar}}^2+H^2}\cdot \sqrt{\left ( D-d_{ar} \right )^2+H^2 }},\\
\nonumber
  &=\frac{{d_{ar}}^2+H^2-d_{ar}D}{\sqrt{{d_{ar}}^2+H^2}\cdot \sqrt{\left ( D-d_{ar} \right )^2+H^2 }}.
\end{align}
Therefore, the incident angle $\theta$ is given by
\begin{align}\label{eq:41}
   \theta = \frac{1}{2}\cdot\arccos{\frac{{d_{ar}}^2+H^2-d_{ar}D}{\sqrt{{d_{ar}}^2+H^2}\cdot \sqrt{\left ( D-d_{ar} \right )^2+H^2 }}}.
\end{align}
Substituting (\ref{eq:39}) and (\ref{eq:41}) into (\ref{eq:38}),  $\beta_r$ can be expressed as a function of $H$ and $d_{ar}$.
This function, however, is complicated, making it difficult to find a closed-form expression of effective $(d_{ar}, H)$. 
As a result, we give some numerical results as follows.  
\begin{figure}[!h]
    \centering
    \includegraphics[width=0.9\linewidth]{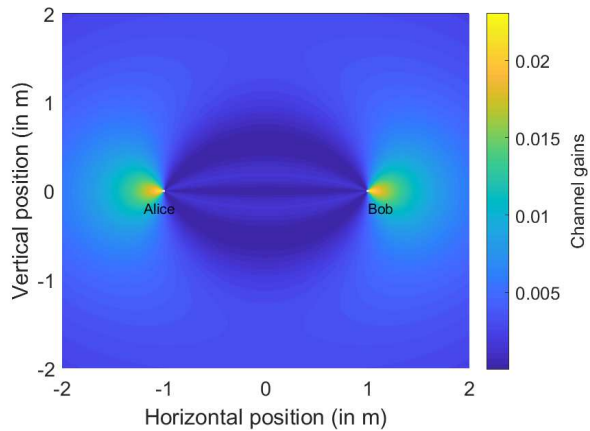}
    \caption{RIS channel gains at different positions.}
    \label{fig:RIS_posigion}
\end{figure}
Fig.~\ref{fig:RIS_posigion} shows the energy gain of the RIS path for different positions of RIS, which is calculated according to  (\ref{eq:38}). As observed, $\beta_r$ is significantly larger than others, when it is placed near Alice/Bob. It is because in this case, $\theta$ approaches zero and $d_r$ is close to $D$, guaranteeing a high reflectivity and a small attenuation, respectively. 

From the numerical results, $\beta_r$ is still larger than $0.01$, when the distance between Alice/Bob and RIS is 20\% of that between Alice and Bob. In other words, it is feasible for Eve to perform RIS-jamming attacks in these regions. For example, in an indoor room, the malicious RIS could be placed on a wall disguised as a painting behind the target terminal. Moreover, in practice, the attack region will be extended by exploiting an active RIS integrated with a power amplifier~\cite{9998527}.

\section{CPR-CRKG: A Countermeasure in wideband OFDM systems} 
\label{sec:detection_and_countermeasure_design}
This section introduces a countermeasure to resist the previously described RIS-jamming attacks and demonstrates its effectiveness through numerical simulations.
\subsection{Protocol Description}
Generally, a RIS-jamming attack is challenging to prevent. 
The key reason for this is the entirely passive operation of the RIS, making the RIS-jamming attack almost imperceptible. 
Further, in view of RIS jamming, traditional anti-jamming techniques such as fast frequency hopping should be considered ineffective. That is, the RIS reflects and alters the legitimate signals regardless of their frequency (assuming a sufficiently large RIS operation bandwidth).

From the previous study of the secret key rate, we have found that although Eve decreases the similarity of channel frequency responses over all subcarriers, only the RIS-induced path is affected.  
If the contaminated path can be distinguished from others, Alice and Bob are still able to generate secret keys from the remaining uncontaminated paths. Following this intuition, we propose a new secret key generation protocol based on contaminated path removal, referred to as CPR-CRKG. Its main task is to resist the above RIS-jamming attacks while still being able to generate a matching key material. {Fig.~\ref{fig:CPR-CRKG} depicts the block diagram of CPR-CRKG and its protocol is elaborated in the following. 
\begin{figure}[h]
    \centering
    \includegraphics[width=0.82\linewidth]{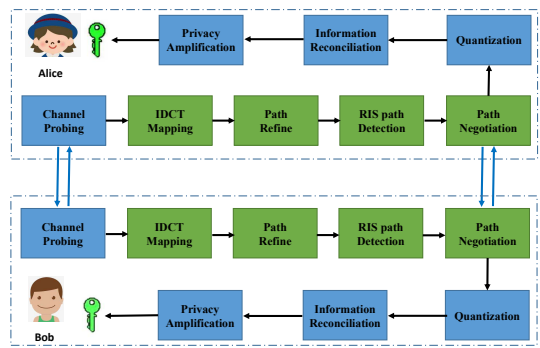}
    \caption{Block diagram of the proposed CPR-CRKG method.}
    \label{fig:CPR-CRKG}
\end{figure}
  \begin{itemize}
        \item [1)] Like in a conventional CRKG procedure, Alice and Bob alternatingly transmit pilot signals to each other. During each channel coherence time, $N_p$ pairs of channel estimates, i.e., 
\begin{align}
  \tilde{\H}_{ba} & = [\tilde{\h}_{ba}^1, \tilde{\h}_{ba}^2, \cdots, \tilde{\h}_{ba}^{N_p}],\\ \nonumber
   \tilde{\H}_{ab} & = [\tilde{\h}_{ab}^1, \tilde{\h}_{ab}^2, \cdots, \tilde{\h}_{ab}^{N_p}],
\end{align} 
are collected by Alice and Bob, respectively.  
  \item [2)] Taking advantage of the wideband OFDM system, Alice and Bob perform an Inverse Discrete Cosine Transformation (IDCT) to map 
$\tilde{\H}_{ba}$ and $\tilde{\H}_{ab}$ into the time-domain, which are obtained by 
\begin{align}
   \tilde{\G}_{ba} =\F  \tilde{\H}_{ba}, \tilde{\G}_{ab} =\F  \tilde{\H}_{ab},
\end{align}
where the transformation matrix $\F \in \Cb^{K \times K}$ is given by
\begin{align}
\F=\left[\begin{matrix} 
&1 &1 &\cdots &1 \\ 
&e^{j2\pi/K} &e^{j4\pi/K} &\cdots &e^{j2K\pi/K} \\ 
&e^{j4\pi/K} &e^{j8\pi/K} &\cdots &e^{j4K\pi/K} \\ 
&\vdots &\vdots &\ddots &\vdots\\
&e^{j2K\pi/K} &e^{j4K\pi/K} &\cdots &e^{j2K^2\pi/K}
 \end{matrix}\right].
\end{align}
Rows and columns of the transformed matrices, including $\tilde{\G}_{ba}$ and $\tilde{\G}_{ab}$, represent relative time delays and channel probing rounds, respectively. 

\item [3)] Since the number of channel paths is far less than $K$, $\tilde{\G}_{ba}$ and $\tilde{\G}_{ab}$ are both sparse matrices that should be refined. For each column of them, the top $N_{sel}$ rows with the highest power gain are regarded as significant paths at a certain channel probing round. To facilitate understanding, we use $d_{k,n} = 1$ to present $\tilde{g}(k,n)$ as a significant path at a relative time delay $k$ and channel probing round $n$, otherwise $d_{k,n} = 0$. To avoid accidental error, row $k$ is recorded in a set $\mathbb{K}_A/\mathbb{K}_B$ by Alice/Bob only when it is recognized as significant paths for no less than $\alpha$ times among all $N_p$ columns, i.e., $\sum_{n=1}^{N_p} d_{k,n} \ge \alpha$. 
%
\item [4)] Alice and Bob detect whether a potential RIS path exists in their significant path set. 
Recalling that for the RIS-jamming attack based on the asynchronism of RIS configurations, Eve needs to increase the switching frequency of the RIS to introduce a considerable temporal deviation. In this case, the channel gain of the RIS-induced path would change more quickly than other paths. Following this idea, we use the autocorrelation coefficients of paths for detection. For each path $k \in \mathbb{K}_A/ \mathbb{K}_B$, Alice and Bob calculate its autocorrelation function as 
\begin{align}
R(j)=\sum_{n=1}^{N_p}\tilde{g}(k,n+j)\tilde{g}^*(k,n),
\end{align}
where $j$ is the offset of the channel probing round. If one path has a significantly more rapidly descending autocorrelation function than others in the set, this path is recognized as a RIS-induced path. Alice and Bob drop their recognized RIS-induced paths from $\mathbb{K}_A$ and $\mathbb{K}_B$, respectively, and would thus mitigate the effect of a potential RIS-jamming attack. 
\item [5)] Alice and Bob negotiate a group of available channel paths by exchanging $\mathbb{K}_A$ and $\mathbb{K}_B$. The intersection set $\mathbb{K} =\mathbb{K}_A \cap \mathbb{K}_B $ is selected as the group of negotiated available channel paths.
For the RIS-jamming attack based on the asymmetry of RIS structures, the RIS-induced path only may appear in $\mathbb{K}_B$ and thus would not be selected as an available channel path. Therefore, the negative effect of this RIS-jamming attack is also avoided. 
Alice and Bob average those channel gains of paths in $\mathbb{K}$ on the column and obtain
\begin{align} 
\tilde{\g}_{ab} =[\tilde{g}_{ab}(1),\tilde{g}_{ab}(2),\cdots,\tilde{g}_{ab}(K_e)]^T,\\ \nonumber
\tilde{\g}_{ba} =[\tilde{g}_{ba}(1),\tilde{g}_{ba}(2),\cdots,\tilde{g}_{ba}(K_e)]^T,
\end{align}
where $K_e$ is the number of paths in  $\mathbb{K}$ and the averaged value on the $k$-th path is given by $\tilde{g}_{ab}(k) =\frac{1}{N_p} \sum_{n=1}^{N_p} \tilde{h}_{ab}^n(k)$.
\item [6)] Alice and Bob collect $\tilde{\g}_{ab}$ and $\tilde{\g}_{ba}$ over multiple channel coherence times and quantize the gain of each channel path using single-bit CDF quantization~\cite{2018trans}. These quantized bit strings, denoted by $\mathbf{k}_{A}$ and $\mathbf{k}_{B}$, are also referred to as raw secret keys, which can be converted into secret keys through information reconciliation and privacy amplification. As these steps are similar to those used in existing key generation schemes, we do not pay particular attention to them and focus our attention only on mitigating RIS-jamming attacks and improving the similarity between $\mathbf{k}_{A}$ and $\mathbf{k}_{B}$.
\end{itemize} 
As outlined above, CPR-CRKG wipes out the contaminated path and therefore removes the effect of the RIS from the channel, effectively mitigating the effect of both asymmetric RIS structures and asynchronous RIS configurations.
The overall computational complexity of CPR-CRKG is $\mathcal{O}(K^2N_p+K\log K+N_{sel}N_p)$.} 
Notably, the proposed CPR-CRKG algorithm can reduce not only the effect of the RIS-jamming attack but also that of the additive noise. Please note that denoising is also necessary for CRKG to achieve good similarity between raw keys and to relieve the burden of information reconciliation. Compared with other denoising schemes \cite{18PCA}, CPR-CRKG only adds an extra RIS path detection step, which involves in additional computational complexity $\mathcal{O}(N_{sel}N_p)$. For this reason, the CPR-CRKG algorithm does not cause notable additional complexity than existing denoising algorithms in the cases of small $N_{sel}$.

\subsection{Numerical Simulation}
We demonstrate the negative effects of RIS-jamming attacks and verify the effectiveness of CPR-CRKG in a wideband OFDM system with the aid of numerical simulations. {As shown in Table. \ref{tb:2}, the system bandwidth is set as 100 MHz and other simulation settings remain the same as that given in Section~\ref{sec:VB}.} 

\begin{table*}
	\centering
	\caption{Simulation setup.}
	\label{tb:2}
	\begin{tabular}{| c | c |}
	\hline
		{Parameters} & {Values} \\ 
		\hline
		{Number of paths $L$}& {4}\\
		\hline
		{Distance $D$} & {$10$ m}  \\ 
		\hline
		{Number of RIS elements} & {$10$} \\ 
		\hline
		{Dielectric substrate of the meta-atom} & {RO4003}\\
		\hline
		{Relative permittivity} & {3.55}\\
		\hline	
		{RIS power $\beta$} & {$1$} \\ 
		\hline	
		{Modulation} & {OFDM} \\
		\hline
		{Number of used subcarriers} & {$128$} \\ 
		\hline
		{System bandwidth} & {100MHz}\\
		\hline
	\end{tabular}
\end{table*}

\begin{figure}[!h]
    \centering
\includegraphics[width=0.8\linewidth]{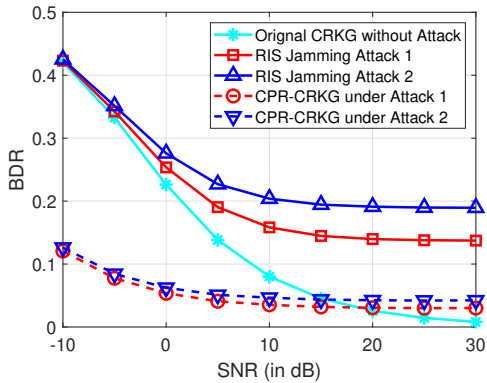}
    \caption{The BDR of different CRKG schemes versus SNR ($\beta = 1$).}
    \label{fig:BDR_SNR}
\end{figure}

Figure~\ref{fig:BDR_SNR} reports the BDR of raw secret keys as a function of SNR. The BDR is defined as the ratio between the number of disagreement bits and the number of total bits. Generally, all BDR curves drop with SNR. The difference is that the BDR of the original scheme is close to zero at high SNR, while the last four schemes have an error floor. Specifically, without the RIS-jamming attack, the BDR of the original CRKG scheme is lower than 0.1, when SNR surpasses 10 dB. 
Under RIS-jamming attacks, the BDR curves fall and reach an error floor of 0.15 and 0.2, respectively. 
The RIS-jamming attack based on the asymmetry of RIS structures (Attack 1) has a lower BDR than that based on the asynchronism of RIS configurations (Attack 2), which is in agreement with the theoretical analysis in Section~\ref{sec:analysis}. The dashed curves of BDR are roughly one-fourth that of their corresponding solid curves, demonstrating the effectiveness of the proposed CPR-CRKG scheme.
CPR-CRKG even shows a lower BDR than the original CRKG scheme without a RIS-jamming attack when the SNR is less than 15 dB.
It is because the proposed CPR-CRKG algorithm can reduce not only the effect of RIS-jamming attack but also the additive noise.
Notably, the BDRs of CPR-CRKG also have error floors, which is caused by the fact that the system bandwidth is limited, so the non-reciprocity caused by the RIS-induced path would also affect other paths after the IDCT mapping step of CPR-CRKG. As attack 2 causes more severe non-reciprocity than attack 1, the BDR of CPR-CRKG under attack 2 is slightly higher than that under attack 1. 

\begin{figure}[!h]
    \centering  \includegraphics[width=0.8\linewidth]{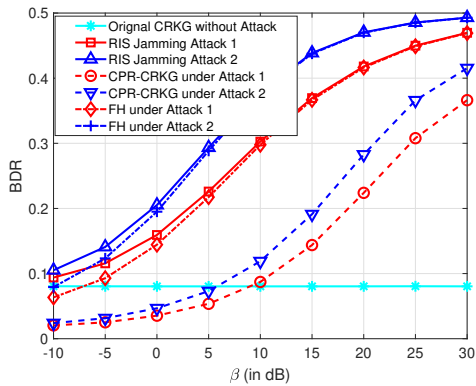}
    \caption{Different BDRs versus $\beta$ (SNR = 10 dB).}
    \label{fig:BDR_Beta}
\end{figure}

Figure~\ref{fig:BDR_Beta} shows the BDRs of different schemes as a function of $\beta$. Except for the original CRKG scheme, the BDR of all other considered schemes rises with $\beta$ rapidly, which is consistent with the results in Fig.~\ref{fig:BDR_Beta}. It is also observed that the gaps between CRKG under attack and our CPR-CRKG scheme 
are enlarged when $\beta$ increases from 0~dB to 20~dB. This is because when $\beta$ is large, the RIS-induced path becomes more evident and is thus easier to detect.
The results in Fig.~\ref{fig:BDR_SNR} and Fig.~\ref{fig:BDR_Beta} substantiate the statement that RIS-jamming attacks have a non-negligible negative effect on the BDR while the proposed CPR-CRKG scheme can mitigate the effect of the RIS-jamming attack. 
Due to the reflection-based operation of RIS over a sufficiently large bandwidth, the jamming signal will inherently follow, regardless of which frequency Alice and Bob hop to. As a result, the BDRs of the FH-based approach are almost the same as those under attacks, much higher than those of the proposed CPR-CRKG approach, which indicates that the RIS-jamming attack can hardly be resolved by FH-based anti-jamming approaches.

\section{Experimental Results} 
\label{sec:experimental_results}
In this section, we present the results of experimental investigations of the RIS-jamming attack based on the asynchronism of RIS configurations and its countermeasure.

\subsection{Experimental Setup}
In our experiments, we place the parties Alice, Bob, and Eve within line-of-sight in an indoor environment. The legitimate parties Alice and Bob perform a CRKG procedure with the goal of establishing a common encryption key. The passive attacker Eve pursues the goal of disrupting the CRKG procedure with a RIS. We now outline additional details of key elements of the setup. {The experiment is orchestrated by a Dell OptiPlex~7090 computer, receiving CSI~data of Alice and Bob via Ethernet and configuring the RIS via USB.}


\Paragraph{Alice and Bob} 
{We utilize a pair of APU~2E4~single-board computers from the manufacturer PC~Engines to represent Alice and Bob. These are based on an AMD GX-412TC processor and have two miniPCI express sockets. We take advantage of the latter to utilize standard IEEE 802.11n \mbox{Wi-Fi} network interface cards~(NICs). The computers run Ubuntu Linux and the open-source ath9k NIC driver in conjunction with the Atheros~CSI tool~\cite{xie_precise_2015}. The parties transmit at approx.~\SI{5}{dBm} on \mbox{Wi-Fi} channel 60 at~\SI{5.3}{\GHz} with \SI{40}{\MHz} bandwidth. For each received packet and spatial MIMO channel, both parties obtain a complex vector containing the CSI data for each of the $114$ non-zero OFDM subcarriers.}

\begin{figure}
    \centering
    \includegraphics{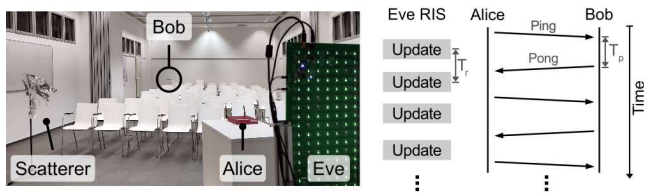}
    \caption{Left: Experimental setup in a seminar room. Right: Illustration of the CRKG channel probing timing and the attacker's concurrent RIS operation.}
    \label{fig:experimental_setup_procedure}
\end{figure}

In our experiments, Alice and Bob exchange \mbox{Wi-Fi} packets in a ping-pong manner to perform bidirectional channel probing, see right side of Figure.~\ref{fig:experimental_setup_procedure}. The channel probing rate $f_p$ is adjustable by delaying the pong response packets by $T_p = \frac{1}{f_p}$ (relative to the reception of ping packets). The time series of channel measurements are translated to bit strings using single-bit CDF quantization~\cite{2018trans}. 
We record $10000$~bidirectional channel measurements, yielding bit sequences of length $10000$~bit. To assess the CRKG performance, we utilize the average BDR over OFDM subcarriers and spatial channels at the output of the quantizer. Please note that we report the raw BDR without applying any information reconciliation and privacy amplification.

\Paragraph{Eve} 
We use a prototype RIS with $256$ unit-cell elements arranged in a $16 \times 16$~array on an FR4 PCB measuring \SI{43}~$\times$~\SI{35}{\cm}. Each rectangular unit cell reflector utilizes a PIN~diode to achieve $1$-bit phase control, i.e., an impinging wave is either reflected with phase shift $0$ or $\pi$. The RIS prototype is designed to realize this behavior at around~\SI{5.35}{\GHz}. {An onboard STM32 microcontroller takes commands via USB and controls the unit cells by means of cascaded shift registers.} The RIS prototype is an evolution of our design from~\cite{heinrichs.2020}, where we used a higher-cost substrate and a smaller array size. Following~Section~\ref{sec:ris_attack_scheme}, in our experiments we apply random RIS configurations with an update period $T_{r} = \frac{1}{f_{r}}$, see right side of Figure.~\ref{fig:experimental_setup_procedure}.

\Paragraph{Environment}
We conduct the experiment in an ordinary indoor environment at our institute building. Since this room does not exhibit sufficient variation for CRKG, we introduce randomness by means of a random scatterer. It consists of a rod holding thin stripes of aluminum foil which move randomly due to airflow from a cooling fan.

\subsection{Timing Aspects}
\begin{figure}
    \centering
    \includegraphics{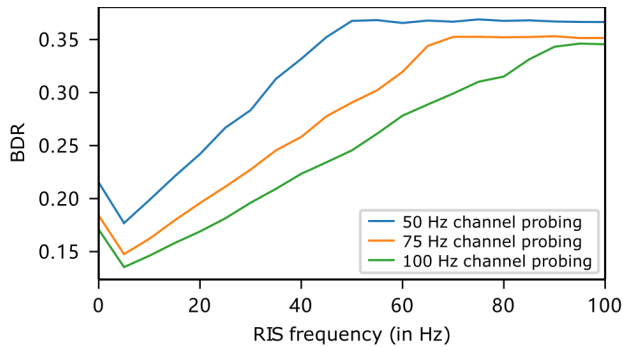}
    \caption{BDR of Alice and Bob over RIS modulation speed.}
    \label{fig:ber_vs_irs_speed}
\end{figure}

For RIS jamming, the attacker aims to disrupt the CRKG channel probing phase by changing the RIS configuration during the bidirectional measurement of the channel. Thus, the attack performance depends on the speed of the adversarial RIS and the legitimate channel probing. To assess this experimentally, we place the parties as shown in Figure~\ref{fig:experimental_setup_procedure}. For various RIS updating and channel probing rates, we take $500$~bidirectional channel samples. Figure~\ref{fig:ber_vs_irs_speed} shows the BDR of Alice and Bob as a function of the RIS modulation frequency for channel probing rates \SI{50}{}, \SI{75}{}, and \SI{100}{\Hz}. 

We can see that the BDR increases linearly with the RIS frequency until a plateau is reached when $f_p = f_{r}$. This is expected since the probability of a RIS update taking place between ping and pong packets is $f_{r} / f_p$ as long as $f_{r} < f_p$ (assuming the RIS reconfiguration and packet duration times to be negligible). Further, we can see that the BDR without an attack at $f_p = 0$ reduces with the channel probing rate as effects from non-simultaneous channel probing are minimized. Interestingly, for low RIS frequencies such as $f_{r} \approx 5~\textrm{Hz}$, the BDR is reduced since the RIS modulation contributes more to the channel variation than corrupting channel reciprocity. 

Our experiment shows that RIS-jamming attacks are feasible in practice by rapidly applying random configurations to the RIS. The attack is of low complexity as the attacker operates independently and asynchronously from the legitimate CRKG process. However, to reach maximum attack efficiency, the RIS modulation speed should be chosen at least as high as the legitimate channel probing speed.

\subsection{Attacker Position}

Next, we investigate how the RIS position affects the attack's effectiveness. From Section~\ref{subsec:secret_key_rate_under_ris_attack}, we expect the ideal RIS position for successful RIS jamming to be in proximity to the legitimate parties with a small incident angle.

\begin{figure}
    \centering
\includegraphics{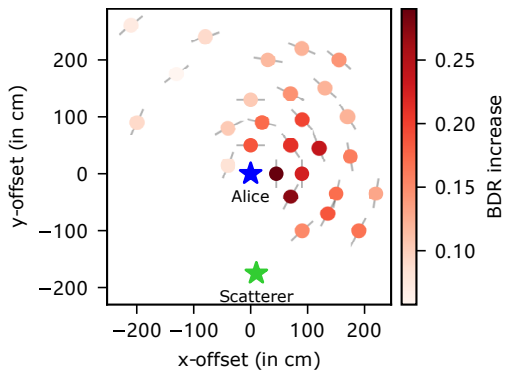}
    \caption{Positions of the attacker RIS. The color scheme indicates the BDR increase due to the RIS-jamming attack. The grey lines indicate the RIS orientation (facing towards Alice). Bob is located at $(-1240~\textrm{cm}, -275~\textrm{cm})$.}
    \label{fig:irs_positions}
\end{figure}

In the experiment, we place the RIS at 27 different positions facing toward Alice. The positions (see Figure~\ref{fig:irs_positions}) were chosen to cover a number of combinations of distances and incident angles. For each position, the RIS first remains constant and then is modulated at \SI{100}{\Hz} update frequency. For both cases, the legitimate parties perform channel probing at~\SI{100}{\Hz} to collect $5000$ bidirectional channel samples. 

We quantify the attack effectiveness using the BDR increase between the cases with modulated and constant RIS. Figure~\ref{fig:irs_positions} shows the BDR increase for each RIS position. We can clearly observe that the attack works best when the attacker is rather close to Alice with a small incident angle which is consistent with the theoretical results. In our experiment, the RIS was capable of increasing the BDR by up to approx.~\SI{30}{\percent} in close proximity to Alice. While such positions may be considered impractical to launch an attack, we still observed reasonable attack performance with BDR~increases of around~\SI{20}{\percent} at positions with approx.~\SI{2}{\m} distance. Despite that, active RIS implementations, e.g.,~\cite{9817032}, which are equipped with power amplifiers could be utilized to overcome distance limitations. 

\subsection{Path Separation}
Next, to assess the feasibility of the CPR-CRKG scheme described in Section~\ref{sec:detection_and_countermeasure_design}, we experimentally evaluate path separation in the time domain as a countermeasure against RIS jamming. To obtain the accurate power delay profile~(PDP) from \mbox{Wi-Fi} CSI data, additional processing is necessary~\cite{xie_precise_2015} to address phase imperfections. Therefore, to simplify the study and solely focus on the proposed countermeasure, we employ a Keysight~P9372A vector network analyzer~(VNA) for channel probing. Other than this, the experimental setup remains as outlined before. We connect the VNA via coaxial cables to one antenna of each party and measure the complex scattering parameters $S_{21}(f)$ and~$S_{12}(f)$. These represent the forward and reverse transmission measured by Alice and Bob, respectively. We measure over an~\SI{80}{\MHz} bandwidth to achieve a spatial resolution comparable with typical \mbox{Wi-Fi} and take $80$~points for each measurement which yields a VNA sweep time of~approx.~\SI{16}{\ms}. 

\begin{figure}
\subfigure[Power over time, $14^{\textrm{th}}$~PDP tap.]{
\centering
\includegraphics[width=.85\linewidth]{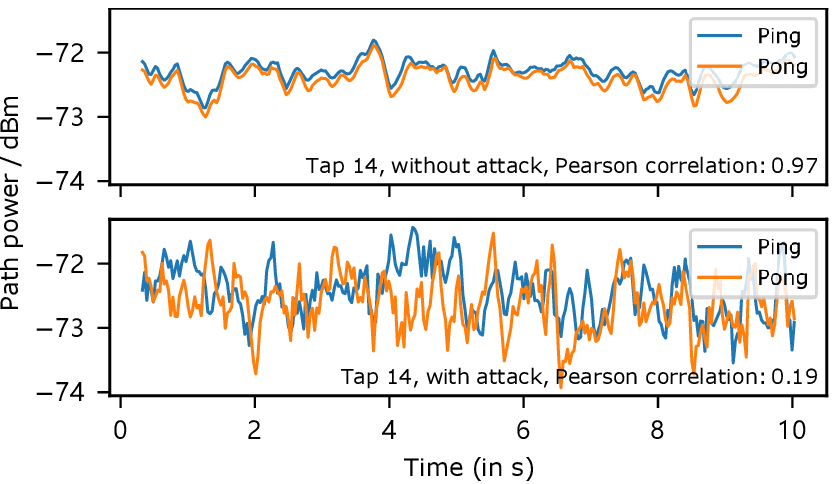}
}
\\
\subfigure[Power over time, $18^{\textrm{th}}$~PDP tap.]{
\centering
\includegraphics[width=.85\linewidth]{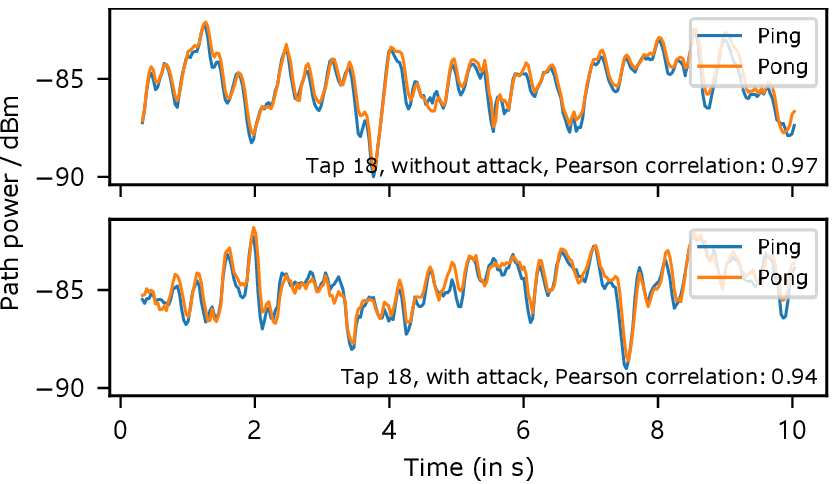}
}
\caption{PDP taps with (upper) and without (bottom) attack.}
\label{fig:ber_pdp}
\end{figure}

For the experiment, we place the RIS in approx.~\SI{1}{\m} distance to Alice. We position the random scatterer such that the difference in path lengths between the RIS and the scatterer is greater than the spatial resolution of approx.~\SI{3.75}{\m}. We then perform channel probing with and without the RIS being randomly configured over time. Using the inverse Fourier transform, we obtain the PDP from the frequency domain measurements. As outlined in Section~\ref{sec:detection_and_countermeasure_design}, it is possible to separate RIS-affected channel paths in the delay domain. To demonstrate the underlying principle, we plot the magnitudes of the $14^{\textrm{th}}$ and $18^{\textrm{th}}$~PDP taps with and without attack in Figure~\ref{fig:ber_pdp}. The propagation paths via the RIS fall into PDP~tap~$14$, causing severe degradation of channel reciprocity as indicated by the Pearson correlation coefficient dropping from $0.97$ to $0.19$. For PDP~tap~$18$, however, the RIS only has minimal effect and channel reciprocity remains intact. {Our experiment demonstrates the feasibility of separating the RIS path, a crucial prerequisite for the successful operation of CPR-CRKG. Therefore, given the ability to isolate the source of randomness and the RIS propagation paths, we believe that CPR-CRKG can serve as an effective countermeasure against RIS jamming in wideband systems. For instance, by employing the attack detection and the protocol outlined in~Section~\ref{sec:detection_and_countermeasure_design}, Alice and Bob can exclude PDP~tap~$14$ and utilize unaffected alternative paths like PDP~tap~$18$ for executing the CRKG procedure. A full implementation and comprehensive evaluation of CPR-CRKG is left for future work.}

\section{Conclusion}
\label{sec:conclusion}
In this paper, we studied CRKG under a new RIS-enabled attack which we coin RIS-jamming attack. In the attack, the reciprocal direct link between the legitimate parties is jammed by the RIS-induced link, effectively reducing reciprocity. A general RIS-involved model was built and on that basis, the principle of RIS-jamming attack was introduced. We elaborated on three examples of the RIS-jamming attack realizations and evaluated them in view of attack requirements, destructiveness, and implementation. 
The attack effect was then studied by formulating the secret key rate with a relationship to the deployment of the malicious RIS. Numerical results verified that RIS-jamming attacks can reduce the secret key rate significantly and the reduction rises with the SNR and the RIS unit's power, which were in agreement with the theoretical analysis. Next, we proposed CPR-CRKG as a countermeasure to resist RIS-jamming attacks. This countermeasure exploits wideband signals for path separation to distinguish the malicious RIS-induced path, deriving secret keys from the remaining channel path gains. We conducted both simulations and experiments on the RIS-jamming attack and CPR-CRKG. The experimental platform consisted of commodity Wi-Fi devices in conjunction with a fabricated RIS prototype. Both simulation and experimental results substantiated the statement that RIS-jamming attacks have a non-negligible negative effect on BDR. Moreover, our results show that the proposed CPR-CRKG scheme mitigates the effect of RIS jamming in wideband systems as long as the source of randomness and the RIS propagation paths are separable. 

Our future work will focus on three aspects. First, we believe that enhancing the attack performance is possible, bearing much potential for future work on using, e.g., an active RIS for amplified signal reflection or optimized instead of random RIS configurations. Second, countermeasures for scenarios where the system bandwidth is not wide enough should be further investigated. Third, exploring the feasibility of locating the RIS would be challenging but valuable.
\bibliographystyle{IEEEtran}
\bibliography{IEEEabrv,citation}

\end{document}